\documentclass[prb,twocolumn,floatfix,showpacs,preprintnumbers,superscriptaddress,amsmath,amssymb]{revtex4}
\usepackage[dvips]{graphics}

\usepackage{mathptmx}
\usepackage{psfrag}
\usepackage{graphicx}

\newcommand{\bra}[1]{\ensuremath{\left\langle#1\right|}}
\newcommand{\ket}[1]{\ensuremath{\left|#1\right\rangle}}

\begin{document}

\title{Transport in metallic multi-island Coulomb blockade systems: A systematic perturbative\\expansion in the junction transparency }

\author{Bj\"orn Kubala}
\affiliation{Institut f\"ur Theoretische Physik III,
Ruhr-Universit\"at Bochum, D-44780 Bochum, Germany}
\author{G\"oran Johansson}
\affiliation{Applied Quantum Physics, MC2, Chalmers, S-412 96 G\"oteborg,
Sweden}
\author{J\"urgen K\"onig}
\affiliation{Institut f\"ur Theoretische Physik III,
Ruhr-Universit\"at Bochum, D-44780 Bochum, Germany}

\date{\today}

\begin{abstract}
We study electronic transport through metallic multi-island Coulomb-blockade 
systems.
Based on a diagrammatic real-time approach, we develop a computer algorithm 
that generates and calculates all transport contributions up to 
second order in the tunnel-coupling strengths for arbitrary multi-island
systems.
This comprises sequential and cotunneling, as well as terms corresponding 
to a renormalization of charging energies and tunneling conductances. 
Multi-island cotunneling processes with energy transfer between different 
island are taken into account.
To illustrate our approach we analyze the current through an island in Coulomb blockade, 
that is electrostatically coupled to a second island through which a large current is flowing. 
In this regime both cotunneling processes involving one island only as well as multi-island processes are important. 
The latter can be understood as photon-assisted sequential tunneling in the blockaded island, where the photons are provided by potential 
fluctuations due to sequential tunneling in the second island. 
We compare results of our approach to a P(E)-theory for photon-assisted 
tunneling in the weak coupling limit.
\end{abstract}


%
%
%
%

\maketitle

\section{Introduction}
Charging effects strongly influence the tunneling of electrons through small metallic, normal-state islands,
 leading to the well-known Coulomb blockade of transport at low temperatures.\cite{FultonDolan87,KulikShekhter75,AverinLikharev86}
Early theoretical and experimental studies explored Coulomb-blockade effects in a single gated island \---a so-called single electron transistor (SET)\---in great detail.\cite{GrabertDevoretBook92}

\begin{figure}[b]
\centerline{\includegraphics[width=0.8\columnwidth]{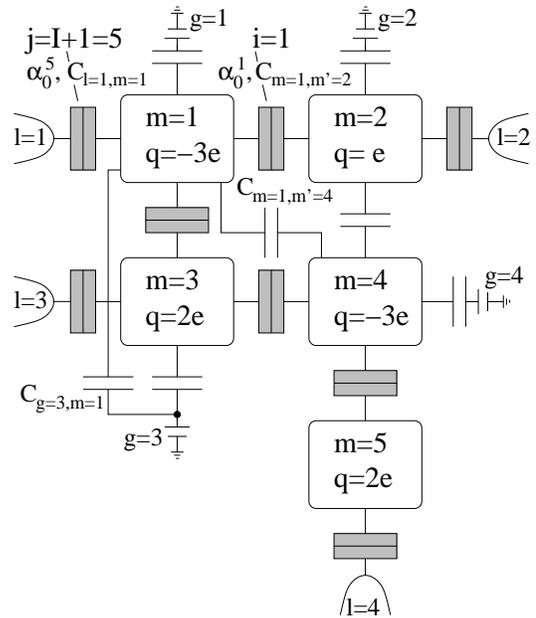}}
\caption{An example of a multi-island geometry. The system
consists of $M=5$ islands with $I=4$ island-island and $J=4$ island-lead
junctions. Any junction, e.g. $i=1$, in between a pair $(m_1=1, m'_1=2)$
of islands is characterized by a transparency $\alpha_0^1$
[see Eq.~(\ref{alpha_0}) below for definition] and a capacitance
$C_{m=1,m'=2}$. Only some stray capacitances are shown in the sketch.
The state of the system is given by the excess charges on the islands:
$\ket{\chi}=\ket{-3,1,2,-3,2}\;$.}
\label{geometry}
\end{figure}

Soon afterwards a number of theoretical\cite{Amman89,AverinKorotkovNazarov91,Korotkov94,TeemuPRB99} and experimental works also considered multi-island 
systems, where several islands are coupled capacitatively or by tunneling 
junctions (see Fig.~\ref{geometry}). 
One approach\cite{DesteveBook92,SchaeferPhysicaE2003,Lehnert03,Turek05} is 
concerned with using one part of the multi-island system as measurement device 
for the residue.  
Serial arrays of islands\cite{DelsingLikharev89,SaclayFirstPumpPhysicaB91,PumpAndTurnstileIEEE91,Pekola_thermometry94,Martinis7PumpErrorsPRB1999,KellerZimmermanStandardReview2003,PTB04,DelsingNature05,SchaeferSeriesPRB2005} have been put 
forward for metrological purposes: for primary 
thermometry\cite{Pekola_thermometry94} and, operated as electron pumps, as a 
standard for both current and capacitance.\cite{Martinis7PumpErrorsPRB1999,KellerZimmermanStandardReview2003,PTB04,DelsingNature05}
For each cycle of periodically changing the gate voltages one electron is 
transfered through the system.
The main operation of these devices is described by first-order perturbation 
theory in the junction conductances, known as orthodox theory for
single-electron tunneling.\cite{Likharev87}
Contributions from second- or higher-order processes, so-called cotunneling,\cite{AverinOdintsovPhysLettA1989,Mqt-tunneling} are limiting the accuracy of the latter devices.\cite{Martinis7PumpErrorsPRB1999,JensenMartinisHighOrderApproxPRB1992}
In other experiments higher-order tunneling contributions are of fundamental 
physical interest. 
For instance, logarithmical temperature dependence of the conductance 
associated with many-channel Kondo correlations, which govern the 
low-temperature transport properties of a single-electron 
transistor,\cite{Matveev91,Joyez97,Wallisser02} could be explained by a 
second-order perturbation expansion at the resonance 
peak.\cite{KoenigSchoellerSchoen97}
For large transparency of the tunnel junctions, first- and second-order
perturbation theory is no longer applicable, and alternative approaches 
suitable to describe transport in the strong tunneling regime such as 
semiclassical approaches,\cite{GolubevZaikin96,GoeppertGrabert98} 
real-time renormalization-group techniques,\cite{KoenigSchoeller98,TeemuPRB99}
path-integral Monte Carlo simulations,\cite{GoeppertHuepperGrabert00} or 
nonperturbative resummations within a real-time diagrammatic 
formulation\cite{SchoellerSchon94} were put forward.

In this paper we describe a real-time diagrammatic approach to transport
through multi-island systems that allows for a systematic perturbation 
expansion in the tunnel-coupling strengths of the tunnel contacts.
In particular, we develop a computer algorithm that generates and calculates
all possible contributions to second-order transport for arbitrary multi-island
systems.
The theory is a generalization of a diagrammatic real-time approach that was 
invented for single-island devices.\cite{SchoellerSchon94}
In contrast to the latter, where a fully analytical treatment of first- plus 
second-order transport is feasible,\cite{KoenigSchoellerSchoen97} the large
number of second-order diagrams for multi-island systems motivates the 
development and use of a computer algorithm. 
The same idea has been used in advanced computer codes for SET networks based 
on orthodox theory.\cite{multi-island-numerics-Fonseca-1995,Simon_by_ChristophWasshuber}

Second-order transport includes cotunneling processes in the Coulomb-blockade
regime, where sequential tunneling is suppressed.
In the standard description of these cotunneling 
processes,\cite{AverinOdintsovPhysLettA1989,Mqt-tunneling}
energy denominators appear that diverge when approaching the onset of 
sequential tunneling.
These divergencies can be removed by replacing the energy denominator with some
 constant\cite{JensenMartinisHighOrderApproxPRB1992}
or by partial resummation of higher-order 
contributions.\cite{LafargeEstevePRB1993}
Besides cotunneling processes, there are other second-order contributions to
transport that become relevant in the regime in which sequential tunneling 
is not suppressed.
They account for the fact that quantum fluctuations due to tunneling give rise
to a renormalization of both the charging energies and the tunnel coupling 
strengths.
This results in  transport contributions that have the functional
form of sequential tunneling but with renormalized system parameters.
An example is the tell-tale characteristics of a many-channel Kondo-effect
at low temperature in a metallic single-electron 
transistor.\cite{Joyez97,Wallisser02,KoenigSchoellerSchoen97} 
Similar renormalization effects are also found in a diagrammatic real-time
description of second-order transport through single-level quantum 
dots.\cite{Thielmann05}
The virtue of the real-time diagrammatic approach employed in this paper lies 
in the fact that the above-mentioned divergencies of energy 
denominators are automatically regularized, and that the renormalization 
effects are taken into account. 
It, therefore, allows for a complete and consistent evaluation of second-order 
transport.

Second-order transport in multi-island systems is qualitatively different from 
that in single-island devices.
In multi-island systems, cotunneling processes in which two different islands
change their charge occupation may occur.
The energy conservation of the total process may be fulfilled by exciting one
island on cost of the other one.
This introduces a coupling between the two islands that may be of importance
in devices in which one island is used as a measurement tool for the charge state
of the other one, as we will discuss in more detail below.
Our theory includes cotunneling involving either one or more than one island.

The outline of the paper is as follows.
First, we present the extension of the real-time diagrammatic theory\cite{SchoellerSchon94} to multi-island systems.
Then, in Sec.~\ref{algorithm}, we present the algorithmic approach to generate
and evaluate all second-order contributions to transport.
This includes a discussion of the applicability range of our perturbation 
expansion.
Further details concerning calculation of diagrams 
and convergence properties of the perturbative expansion are included in 
Appendixes \ref{calculatingdiagrams} and \ref{convergence}. 
Afterwards, we illustrate our theory by applying it to one of the simplest 
multi-island setup, namely that of two single-electron transistors put in
parallel, Sec.~\ref{example}.
For this example, we discuss the physics of energy exchange between the two
transistors, interpret it within a simplified picture using a 
$P(E)$-description of photon-assisted tunneling, and demonstrate the virtue 
and the limits of this picture by comparing it with our full second-order
calculation.

\section{Real-time diagrammatics for metallic multi-island systems\label{RT-diagrammatics}}

\subsection{Metallic multi-island geometries\label{setup}}
The system we consider in this paper consists of a number $M$ of
small metallic islands as well as $L$ leads, which are connected
to some islands by $J$ junctions (see Fig. \ref{geometry}). Additionally there are $I$
junctions between the metallic islands. Accordingly with any
junction $i=1,\;2, \hdots, I$ we will associate the pair of
islands $(m_i,\; m'_i)$ it is connecting, where the order is
arbitrary but fixed, thus defining a `direction' for the junction.
Likewise for any of the junctions between leads and islands,\cite{footnote1}
indexed by $j=I+1,\; I+2, \hdots,\; I+J$, we define a pair $(l_j,\; m_j)$.
Furthermore we have to take into account the capacitances $C_{lm}$
between islands and leads, $C_{mm'}$ in between two islands, as
well as capacitances $C_{gm}$ to additional external gates $g$.

For typical samples of metallic islands the level spectrum is
dense and the total charge on any island is large. Therefore not
only the leads but also the islands can be considered as large
equilibrium reservoirs, described by Fermi distribution functions.
These equilibrium distribution functions are not influenced by
tunneling processes comprising only a very small fraction of the
overall number of electrons. 
Electron-hole excitations left behind after tunneling are quickly equilibrated\---on a time scale short compared to typical times between tunneling events.

The state of our system is then
described by the excess charges $\chi=\{n_1,\hdots,\; n_M\}$
(total charge minus background charge) sitting on the $M$ islands.

The system can be modeled by the Hamiltonian:
\begin{equation}
\label{Hamiltonian}
H=H_L + H_M + V + H_T = H_0 + H_T .
\end{equation}

The noninteracting electrons in the leads and islands are described by
\begin{equation}
H_L= \sum_{l=1}^L \sum_{\kappa\nu}\tilde{\epsilon}^l_{\kappa\nu}a^{\dagger}_{l\;\kappa \nu}a_{l\; \kappa \nu} , \quad
H_M= \sum_{m=1}^M \sum_{\lambda\nu}\varepsilon^m_{\lambda\nu}c^{\dagger}_{m\;\lambda\nu}c_{m\;\lambda \nu} ,
\end{equation}
where wave vectors $\kappa$ and $\lambda$ numerate electron states
within a given transverse channel $\nu$. (Note, that the
transverse channel index $\nu$ includes the spin of the electrons.
For ease of notation we omit subindices $\kappa_{l\nu}$ or
$\lambda_{m\nu}$ throughout.)

Coulomb interaction of electrons is captured by $V =
V(\hat{\chi})$, the electrostatic energy of a given charge state
$\ket{\chi}=\ket{n_1,n_2,\hdots,n_M}$. It depends in a complex way
on gate and bias voltages and the resulting charges on the
capacitances $C_{lm}$, $C_{mm'}$, and $C_{gm}$, respectively. A
straightforward scheme to calculate this dependence, governed by
classical electrostatics, is given in Ref. \onlinecite{vanderWielReview2003}.
There a capacitance matrix is introduced to first calculate voltages on the 
islands and subsequently the electrostatic energy. For an example of the 
gate and bias voltage dependence of the electrostatic energy, see the discussion 
of a two-island setup in Sec.~\ref{two-island_setup}.

Finally charge transfer through the junctions is depicted by the
tunneling Hamiltonian
\parbox{0.8\columnwidth}{
\begin{eqnarray*}
\label{H_T}
H_T &=&\phantom{+} \sum_{j}\sum_{\kappa\lambda\nu}\left(T^{j\nu}_{\kappa\lambda}a^{\dagger}_{l_j\kappa\nu}c_{m_j\lambda\nu}e^{-i\hat{\phi}_{m_j}} + {\rm c.c.} \right) \\
    &&+     \sum_{i}\sum_{\lambda\lambda'\nu}\left(\mathsf{T}^{i\nu}_{\lambda\lambda'}c^{\dagger}_{m_i\lambda\nu}c_{m'_i\lambda'\nu}e^{-i(\hat{\phi}_{m'_{i}}-\hat{\phi}_{m_{i}})} + {\rm c.c.} \right)
,
\end{eqnarray*}
}\hfill\parbox{0.2\columnwidth}{\begin{eqnarray}\end{eqnarray}}
where $T^{j\nu}_{\kappa\lambda}$ and
$\mathsf{T}^{i\nu}_{\lambda\lambda'}$ are tunneling matrix
elements for junctions $j$and $i$, respectively, and $\exp{(\pm
i\hat{\phi}_{m_{j/i}})}$ is a charge shift operator, acting on the
charge state $\ket{\chi}$ described above.\cite{footnote2}

Involving a phase operator $\hat{\phi}_{m}$ as a canonical conjugate to
the charge operator $\hat{N}_m$ of the island $m$, i.e., $ \left[
\hat{\phi}_m, \hat{N}_m \right] = i$, these operators $\exp{(\pm
i\hat{\phi}_m)}$ changes the excess particle number on the island
$m$ as $n_m \rightarrow n_m\pm 1$ for each tunneling process
accordingly.

\subsection{Diagrammatic technique}
In this section we generalize the diagrammatic technique developed in 
Ref.~\onlinecite{SchoellerSchon94} for a single SET to study multi-islands 
systems as described by Eqs.~(\ref{Hamiltonian})--(\ref{H_T}) above. 
A short overview over the derivation is
given, while the reader is referred to Refs.~\onlinecite{SchoellerSchon94,
KoenigSchoellerSchoen97}, and \onlinecite{SchoellerCuracao1997} for more details. In
Sec.~\ref{singlemulti} we explicitly discuss differences between single and
multi-island case.

\subsubsection{The current}
One of the central objects of our theoretical description is the
probability $P_{\chi}$, to find the multi-island system in a
certain state $\ket{\chi}=\ket{n_1,n_2,\hdots,n_M}$ in charge
space. Experimentally accessible quantities such as the average
charge of a certain island follow directly from this probability.
The other quantity, we will mainly be focused on, is the current
flowing through junction $j$ into reservoir $l_j$, given by the
change in the number of particles,\\
\parbox{0.8\columnwidth}{
\begin{eqnarray*}
I_l(t) &=&e\frac{d}{dt}  \langle \hat{N}_{l_j}(t) \rangle\\
&=& 
e
 \sum_{\kappa\lambda\nu} \left[   T^{j\nu}_{\kappa\lambda}  \langle (a^{\dagger}_{l_j\kappa\nu}c_{m_j\lambda\nu}\;e^{-i\hat{\phi}_{m_j}} )(t)
+ {\rm c.c.} \rangle \right].
\end{eqnarray*}
}\hfill\parbox{0.2\columnwidth}{\begin{eqnarray}\label{current_operator}\end{eqnarray}}
Note that the current operator has a similar structure as the
tunneling Hamiltonian $H_T$.

\subsubsection{Time evolution of operators}
\begin{figure}[h]
\centerline{\includegraphics[width=0.95\columnwidth]{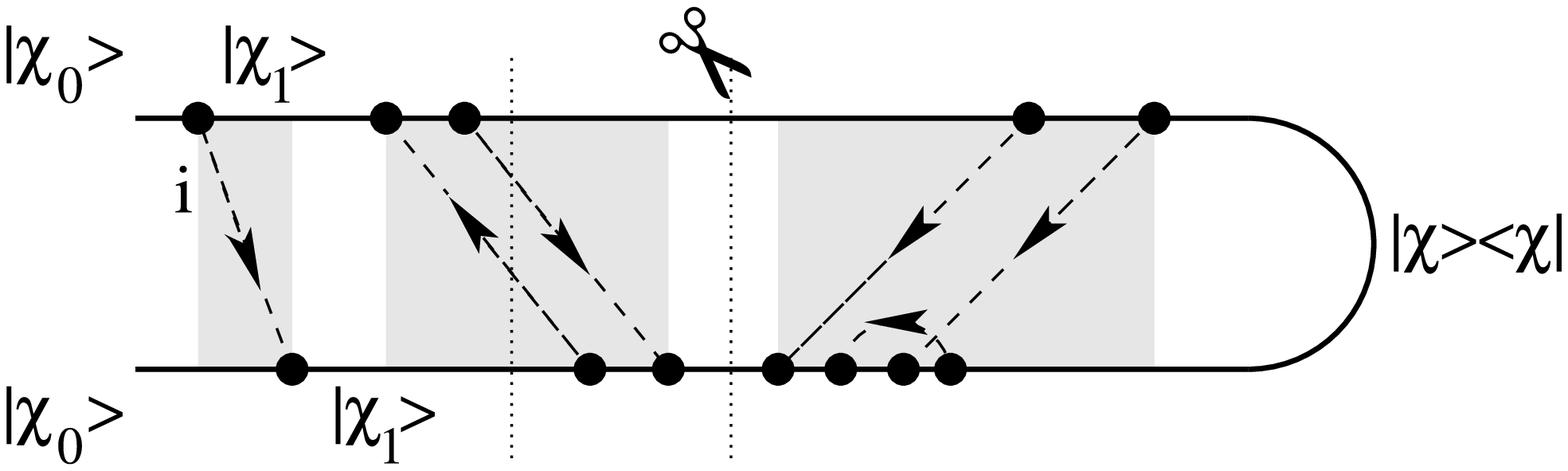}}
\caption{Diagram describing the time evolution of the islands' density matrix with sequential tunneling, cotunneling (two tunneling lines at a time), and higher order processes (from left to right). The diagram is reducible, i.e., it can be cut at a certain time without cutting a tunneling contraction line. Summing irreducible parts (shaded) yields the self-energy. In the left-most part a tunneling process through junction $i$ in between a pair of islands $(m_i, m'_i)$ changes the state of the island on the upper branch from $\ket{\chi_0}=|n_1,n_2,\hdots,n_{m_i},\hdots,n_{m'_i},\hdots,n_M\rangle$ to $\ket{\chi_1}=|n_1,n_2,\hdots,{n_{m_i}-1},\hdots,n_{m'_i}+1,\hdots,n_M\rangle\;$.}
\label{keldysh1}
\end{figure}

The nonequilibrium time evolution of the charge degrees of freedom
is described by the expectation value of the (diagonal)
density matrix
\begin{eqnarray}
\label{time_evolution}
P_{\chi}(t)     &=& \langle\; \left( \ket{\chi}\bra{\chi}\right) (t)  \;\rangle \nonumber\\
        &=& {\rm Tr} \left( \rho_0  T^{+} e^{i \int_{t_0}^{t} dt' H_T(t')_I} \ket{\chi}\bra{\chi} T^{-} e^{-i \int_{t_0}^{t} dt' H_T(t')_I} \right) \nonumber\\
        &=& {\rm Tr}\; \Bigg( \rho_0  \sum_{s=0}^\infty (-i)^s \int\limits_{_K }dt'_1 \!\!\!\!\!\!\int\limits_{_K \atop t'_1<t'_2<\hdots<t'_s  }\!\!\!\!\!\!\!\!\!\!\!\!\!dt'_2 \;\hdots \;\int\limits_{_K}dt'_s  \nonumber\\
&&  \times T^K
\left[ H_T(t'_1)_I H_T(t'_2)_I
\hdots  H_T(t'_s)_I   \ket{\chi}\bra{\chi}\;\right]
\Bigg).
\end{eqnarray}
Here we replace time-ordering $T^+$ and anti-time-ordering
operators $T^-$ by introducing integration $\int_{_K}dt'$ along
the Keldysh contour with ``times'' $t'$ running forward from $t_0$
to $t$ and backward from $t$ to $t_0$ (see Fig.~\ref{keldysh1}). The ordering
of times $t'_1<t'_2<\hdots<t'_s\;$ is with respect to this
Keldysh contour with the Keldysh time-ordering operator $T^K$
arranging the operators in the tunneling Hamiltonian in proper
order. $H_T(t)_I$ denotes the tunneling part of the Hamiltonian [Eq.~(\ref{Hamiltonian})] in 
interaction representation with respect to $H_0$.

Due to the separation of fermionic and charge degrees of freedom
the Hamiltonian $H_0$, including interaction through the charging
energy $V(\chi)$, is bilinear only in electron operators. Thus we can
apply Wick's theorem and perform the trace over these degrees of
freedom by contracting tunneling vertices in pairs. 

In fact, there are two independent contraction lines for each vertex, as each of
the two electron operators $a^{\dagger}_{l_j\kappa\nu}/
a_{l_j\kappa\nu}\;\text{and}\; c^{\dagger}_{m_j\lambda\nu}/c_{m_j\lambda\nu}$
constituting a tunneling vertex is connected to another matching
operator $a_{l_j\kappa\nu}/a^{\dagger}_{l_j\kappa\nu} \;\text{and}\; 
c_{m_j\lambda\nu}/c^{\dagger}_{m_j\lambda\nu}$, not necessarily
both of these being part of the same tunneling vertex. However in
the limit of large transverse channel number ``simple loop''
configurations dominate and we can represent contractions by one
(directed\cite{footnote4})
 double tunneling line between pairs of vertices.
Readers may refer to Figs.~2 and 10 of Ref.~\onlinecite{SchoellerSchon94} for an illustration.

A realization of a resulting diagram is shown in Fig.~\ref{keldysh1}, where
certain physical processes can be identified with parts of the
diagram. The left-most part shows explicitly how the charge state
is changed by a tunneling process across junction $j$. Processes
with several tunneling lines at a time $t$ correspond to higher-order processes such as cotunneling.

\subsubsection{Generalized master equation}
From these diagrams for the time evolution of the density matrix a
formally exact master equation can be derived,
\begin{equation}
\label{mastereqn}
\frac{d}{dt}P_{\chi}(t) = \sum_{\chi' \ne \chi} \int_{t_0}^t
dt' \left[ P_{\chi'}(t')
\Sigma_{\chi',\chi}(t',t) -  P_{\chi}(t')
\Sigma_{\chi,\chi'}(t',t) \right] ,
\end{equation}
where the central object needed as input is the full quantum
mechanical transition rate $\Sigma_{\chi',\chi}(t',t)$ from a
state ${\chi'}$ at time $t'$ to state ${\chi}$ at time $t$. This
rate is the sum over all irreducible (see Fig.~\ref{keldysh1}) diagrams and corresponds to
the self-energy of a Dyson equation for the full propagator of the
system. After Laplace transforming, the stationary distribution
$P_{\chi}$ of charge states is found by
\begin{equation}
\label{probabilities_Laplace}
0=\sum_{\chi'} P_{\chi'} \Sigma_{\chi',\chi} ,\;{\rm with}\;\; \Sigma_{\chi',\chi}= i \int_{-\infty}^0
dt' \;\Sigma_{\chi',\chi}(t',0).
\end{equation}

In a similar manner as for the density matrix, a diagrammatic representation for the expectation value of the current is found. The current operator  becomes manifest in any such diagram as an additional external vertex [cf.~Eqs.~(\ref{H_T}) and (\ref{current_operator})] at the (right-most) time $t$. As above \emph{partial} self-energies can be defined
\begin{equation}
\label{self-energies}
\Sigma_{\chi',\chi}(t',t) = \sum_{j} \left\{ \Sigma^{j+}_{\chi',\chi}(t',t) + \Sigma^{j-}_{\chi',\chi}(t',t)  \right\} .
\end{equation}
Here  $\Sigma^{j+}_{\chi',\chi}$ includes all those diagrams contributing to $\Sigma_{\chi',\chi}$ with the
rightmost vertex on the upper propagator and right-most contraction
line describing tunneling out of lead $l_j$ through junction $j$
as well as diagrams with right-most vertex on the lower propagator
and the right-most process describing tunneling into lead $l_j\;$;
correspondingly summing the remaining self-energy diagrams with right-most tunneling through junction $j$ defines
$\Sigma^{j-}_{\chi',\chi}\;$.

The stationary current then follows as
\begin{equation}
\label{current}
I_{l_j}= -ie\sum_{\chi,\chi'}P_{\chi'} \Sigma^{j+}_{\chi',\chi} = ie\sum_{\chi,\chi'}P_{\chi'} \Sigma^{j-}_{\chi',\chi}.
\end{equation}

\subsubsection{Diagrammatic rules}
Identification of terms of the sums in Eq.~(\ref{time_evolution}) with diagrams
leads to a set of rules (for details see Ref.~\onlinecite{SchoellerSchon94}) for calculating the
value of a certain realization of a diagram. Here we will give
these rules in the form most convenient for calculating the
Laplace transform of the self-energy, 
which allows calculation of stationary probabilities and current via
Eqs.~(\ref{probabilities_Laplace}) and (\ref{current}):
\paragraph*{\rm(1)}
Draw all topologically different diagrams with tunneling
lines and choose junction and direction for each line. Assign
energies $V(\chi)$ to propagators, $\omega_r$ to tunneling lines.
\paragraph*{\rm(2)}
Each segment from $t_s \le t \le t_{s+1}$ gives a factor $[ \Delta
E_s +i\eta ]^{-1}$, where $\Delta E_s$ is difference of left-going
minus right-going energies (including the energies $\omega_r$ of
tunneling lines).
\paragraph*{\rm(3)}
Each tunneling line gives a rate function
$\alpha^{j/i\pm}(\omega_r)$\---as defined below\---for the direction of the tunneling
line across junction $j/i$ going backward/forward with respect to
the Keldysh contour.
\paragraph*{\rm(4)}
There is a prefactor $(-1)$ for each internal vertex on the backward propagator.
\paragraph*{\rm(5)}
Integrate over the energies $\omega_r$ of tunneling lines. \vspace{0.05cm}\\

The golden-rule rate $\alpha^{j\pm}$
($\alpha^{i\pm}$ correspondingly) stems from implicit integration
over the energy of one of the double tunneling lines
\begin{equation}
\label{rate_definition}
\alpha^{j\pm}(\omega)=\int dE \alpha^{j}_0 \; f_{l_j}^\pm(E+\omega)f_{m_j}^\mp(E)
= \pm \alpha^{j}_0 \frac{\omega-\mu_{l_j}}{e^{\pm  \beta (\omega-\mu_{l_j})} -1} ,
\end{equation}
where $f_{l/m}^+$ is the Fermi distribution function of lead $l$
or island $m$ and $f_m^-=1-f_m^+\;$.
Hereby assuming constant tunneling matrix elements
$T^{j\nu}=T^{j\nu}_{\kappa\lambda}$ as well as a constant temperature $k_BT=1/\beta$ across the sample and neglecting energy
dependencies of the density of states $N^{l\!/\!m\;\nu}(E)$, each
junction $j$ is characterized by a single parameter,
\begin{equation}
\label{alpha_0}
\alpha^{j}_0=\sum_{\nu} \left|T^{j\nu}\right|^2 N^{m_j\nu}(0) N^{l_j\nu}(0)=\frac{h}{4\pi^2e^2}\frac{1}{R_{j}}=\frac{R_K}{4\pi^2R_{j}} ,
\end{equation}
related to the tunneling resistance $R_{j}$ of the junction.

\subsubsection{Diagrams for single and multiple islands\label{singlemulti}}
The formalism of real-time diagrammatics as presented in the
previous section is applicable to the study of arbitrary
multi-island system, as defined by the Hamiltonian Eqs.~(\ref{Hamiltonian})--(\ref{H_T}).
Hitherto derivation and application was mainly concerned with the case
of a single island (see, e.g., Ref.~\onlinecite{SchoellerCuracao1997} and references therein) and one
study of two islands in series.\cite{TeemuPRB99} Therefore this
section is devoted to a brief summation of differences between
the single and multi-island cases.

Virtually the entire formalism of real-time diagrammatics
translates from the single to multi-island case by switching over from
one variable of interest, the charge on the island $n$, to the
$M$-tuple $\chi={n_1,n_2,\hdots,n_M}$ of charges on all the $M$
islands. This extension of state space has nontrivial effects
only for the practical application of the method, but not for
general scheme and derivation. 
Complications for the multi-island as compared to the single-island case
arise due to the non-trivial electrostatic charging energies,
numerous tunneling processes mediating between any two charge
states, and most importantly the exponentially increasing number of charge 
states as the number of islands grows.

\section{Algorithmic diagrammatics\label{algorithm}}
As for a general multi-island system a closed analytic solution seems 
unattainable, we set up and use in the following an automatized, 
computer-based numerical approach.

\subsection{Algorithmic scheme}
From our discussion above we learned about the essential steps in
our problem: Solving the electrostatics delivers the energy terms,
needed for calculating rates of certain processes, which allow us to
find stationary solutions of the rate equations. In this section
we present the automatization of this process, where the (by far)
most intricate part is the actual calculation of rates. This
includes the automatic generation and calculation of real-time
diagrams.

The first step in our scheme is the restriction to a finite charge
state space. This subspace will include all states within a few
$k_BT$ around the classical ground state and additionally all
states reachable by simple tunneling events from these classically
occupied states. Note that this choice can be self-consistently
checked from the resulting occupation probabilities. For all these
states the electrostatic energies $V(\chi)$, which also depend on
gate and bias voltages, are calculated according to the method presented in 
Ref.~\onlinecite{vanderWielReview2003}.

\subsection{Transition rates}
We gain the rates entering Eqs.~(\ref{probabilities_Laplace}) and
(\ref{current}) up to second order in
tunneling by generating and calculating all self-energies diagrams
with one or two contraction lines. The scheme of generating all
diagrams of a certain order is presented in the following for
second-order diagrams, first order being trivial, whereas
higher-order diagrams, although easily generated, are considerably
harder to calculate and will not be considered below.

\paragraph*{Generating diagrams:}
Starting with a certain state $\chi'$ of our chosen subspace of
charge states for both, upper and lower branch of the
Keldysh contour (see Ref.~\onlinecite{SchoellerSchon94} for a discussion of the
diagonality of $\Sigma$), the following steps were implemented.

A \emph{first} (i.e., left-most) vertex $v_1$ is placed on either the 
upper or lower branch, a junction $j_1/i_1$ as well as a direction
for the tunneling event is chosen. This determines the charge
state\cite{footnote5}
on both branches in between vertices $v_1$
and $v_2$ and consequently the energy terms determining the ``free''
propagation, governed by $H_0$ and the tunneling line energy
$\omega_1$. For the \emph{second} vertex $v_2$ the same choices
can be independently made, as it is not contracted with the first
vertex. (Such a connection would result in a diagram, reducible to
two first-order blocks, thus not being part of the irreducible
second-order self-energy.) The \emph{third} vertex, however, is
connected to either one of the vertices $v_1$ or $v_2$, inheriting
whereby junction and direction of the tunneling event, whereas
position on upper or lower branch is free to choose. This is also
the only freedom of choice left for the last vertex, which is
combined with the remaining tunneling line. Just as at the first
vertex, charge states have to be changed according to the
tunneling processes at the vertices all along the propagator up to
the right-most final state $\chi$, which is the same on upper and
lower propagator.

Estimating the number of diagrams allows some gauge of the
complexity of the problem. For each charge state (for low
temperatures three per island, resulting in a total of $3^M$)
there are $2^4 \times [2(J+I)]^2 \times 2$ second-order diagrams.
Here the first factor stands for the upper/lower branch for each
vertex, the second results from vertices $v_1$ and $v_2$ for
choosing junction and direction of tunneling, while the last
factor stems from combining vertices into pairs.

For each of these diagrams an analytical integral expression is
immediately given by the diagrammatic rules, discussed above.
Evaluation of these expressions to a numerical value is discussed
in Appendix \ref{calculatingdiagrams}.

\subsection{Stationary state solutions}
The value of any first- or second-order diagram is then added to the
appropriate matrix element of the self-energy matrix
$\Sigma^{(1/2)}_{\chi',\chi}$, where indices ${\chi',\chi}$ are
initial and final charge state of the diagram.

All diagrams required for calculating currents are already created
within this scheme by identifying the right-most tunneling vertex
with an external current operator. Accordingly adding (with the
proper sign) the values of diagrams, where the last vertex involves
junction $j$, yields $\Sigma^{j\pm(1/2)}_{\chi',\chi}$.

$\Sigma^{(1/2)}_{\chi',\chi}$ constitute the first terms of an
expansion of the self-energy in powers of $\alpha_0$,
\begin{equation}
\Sigma_{\chi',\chi} = \sum_{k=1}^\infty \Sigma^{(k)}_{\chi',\chi} .
\end{equation}
Expanding the probabilities $P_{\chi}=\sum_{k=0}^\infty
P^{(k)}_{\chi}$ correspondingly, we find from the stationary rate
equation in first and second order,
\begin{equation}
0 = \sum_{\chi'}P^{(0)}_{\chi'} \Sigma^{(1)}_{\chi',\chi} \;\;{\rm and}\;\; 0 = \sum_{\chi'}P^{(0)}_{\chi'} \Sigma^{(2)}_{\chi',\chi}
+  \sum_{\chi'}P^{(1)}_{\chi'} \Sigma^{(1)}_{\chi',\chi},
\end{equation}
which finally gives solutions for $P^{(0)}_{\chi'}$ and
$P^{(1)}_{\chi'}$. For first- and second-order current\cite{footnote6}
, we likewise find
\begin{eqnarray}
I^{(1)}_j &=& -ie \sum_{\chi,\chi'}P^{(0)}_{\chi'} \Sigma^{j+(1)}_{\chi',\chi} \\
I^{(2)}_j &=& -ie \sum_{\chi,\chi'}\left( P^{(0)}_{\chi'} \Sigma^{j+(2)}_{\chi',\chi}
+  P^{(1)}_{\chi'} \Sigma^{j+(1)}_{\chi',\chi}\right) .
\end{eqnarray}
Thus we derived a scheme to calculate experimental accessible quantities for an arbitrary 
multi-island geometry, namely the average current in any lead.

\subsection{Limits of the approach}

There are both practical and fundamental limits of the approach presented 
above.
As the number of charge states, and consequently the number of diagrams 
to calculate grows rapidly with the number of islands, a straightforward 
simulation of some interesting existing experimental 
applications\cite{Martinis7PumpErrorsPRB1999,Pekola_thermometry} using long 
arrays of islands ($7$\---$100$ islands) is practically infeasible.
Similarly, including higher than second-order contributions would involve
more complex integral expressions that complicates the numerics considerably.

The fundamental limit deals with the question of convergence of the 
perturbative expansion.
As we derive in Appendix \ref{convergence}, the inequality
$\max\{k_BT, \left| \Delta-\mu_j \right|\} > \left|\sigma(\Delta)\right|$
has to be fulfilled for tunneling across any junction $j$. Here $\left| \Delta-\mu_j \right|$ is the 
distance to resonance for a tunneling process across junction $j$ with chemical potential difference $\mu_j$ and
$\sigma(\Delta)$ is the self-energy, characterizing the spectral density of the level into which tunneling occurs at energy $\Delta$.
This incorporates renormalization of level position as well as a finite lifetime width, as reflected in real 
and imaginary part of the self-energy. We elaborate on this statement in Appendix \ref{convergence}, where we 
analyze the example of a single SET by comparison to a nonperturbative approach.

\section{An example: two parallel single-electron transistors\label{example}}
To give an example we will now apply our method to calculate
transport through two single-electron transistors (SETs), where
the two islands are capacitively coupled to each other, see
Fig.~\ref{TwoIslands}. This system was studied experimentally, e.g., in Refs.~\onlinecite{SchaeferPhysicaE2003} and \onlinecite{SCtwoislands}. The capacitive
interaction implies that a change of the charge state of one SET
will change the effective gate voltage of the other, which will
lead among other things to a broadening of the charge state
transitions as described in
Ref.~\onlinecite{SchaeferPhysicaE2003}. This is an effect visible
already in first order of the tunneling conductances. The
second-order contributions include cotunneling in each SET
separately, but also second-order processes in which both SETs are 
involved.
For the latter, the energy gained in a tunneling process in one SET
can be used to excite the other SET to a charge state not accessible for 
first-order transitions.
This results effectively in an energy exchange between the two islands,
a feature that is qualitatively new as compared to the single-island case 
or first-order transport in multi-island systems.
We concentrate our analysis on a regime, where it is this energy exchange, that enables 
transport in an otherwise blockaded region.
One may try to simulate this energy transfer in a simplified picture 
of photon-assisted first-order tunneling in the second SET coupled to a
energy-providing photon bath that represents electrostatic fluctuations by
tunneling processes in the first SET.
This simplified approach and the effect of the energy-transfer processes on 
the transport characteristics will be described in more detail below.
The advantage of this approach is its feasibility and compact analytical
results. 
The disadvantage is that it relies on several approximations, and its 
applicability range is unclear.
This question can be answered, though, by our full-fledged second-order 
transport calculation. 
By comparing the results of both approaches we are able to define the range
of parameters beyond which the simplified picture loses its reliability.

\subsection{Setup\label{two-island_setup}}
\begin{figure}[]
\centerline{\includegraphics[width=6cm]{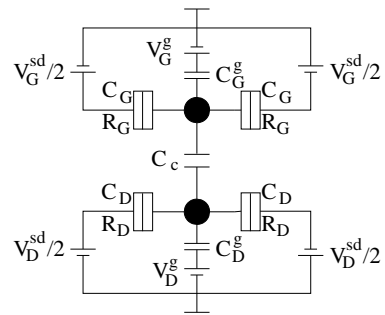}}
\caption{Two capacitively coupled SETs. The upper one is the noise
generator, and the lower the quantum noise detector.}
\label{TwoIslands}
\end{figure}

The system consists of two metallic islands (called noise {\em
generator} and {\em detector} as explained below), each connected
to a separate source and drain lead, see Fig.~\ref{TwoIslands}.
Each junction is generally described by a capacitance and a
tunnel resistance, as discussed in Sec.~\ref{RT-diagrammatics}. For simplicity we
here assume that for each island the connection to source and
drain is symmetric, and also that the source-drain voltage is
applied symmetrically. Note, that we change our notation compared to the general 
setup in Sec.~\ref{setup} in order to make the association with the noise
{\em generating} and {\em detecting} part more transparent. We denote the 
generator/detector junction capacitances $C_G/C_D$, the resistances 
$R_G/R_D$, and the applied source-drain voltages 
$V_G^{\text{sd}}/V_D^{\text{sd}}$. Furthermore the working
point of the generator/detector SET is controlled by gate voltages
($V_G^g/V_D^g$) applied across the gate capacitances
($C_G^g/C_D^g$). The two islands are coupled through the coupling
capacitance $C_c$.

Following Refs.~\onlinecite{vanderWielReview2003} and
\onlinecite{SchaeferPhysicaE2003} we find the electrostatic energy
of our system. This is most straightforwardly described
introducing the sum capacitances of the two islands
\[ C_{G/D}^\Sigma=2 C_{G/D}+C_{G/D}^g+C_c, \]
the rescaled charging energies
\[E_G=\frac{e^2}{2\left(C_G^\Sigma - C_c^2/C_D^\Sigma\right)},
\ \ E_D=\frac{e^2 }{2\left(C_D^\Sigma - C_c^2/C_G^\Sigma\right)},
\] the interaction energy
\[ E_{\text{int}}=\frac{e^2C_c}{C_G^\Sigma C_D^\Sigma - C_c^2}, \]
and also the dimensionless gate charges $n_{x}^{G/D}=C_{G/D}^g
V_{G/D}^g/e$. 
Then the electrostatic part of the Hamiltonian reads
\begin{eqnarray}
V(n_G,n_D) &=& \phantom{+} E_G (n_G-n_x^G)^2 + E_D (n_D-n_x^D)^2 
\nonumber\\
&&+{}E_{\text{int}} (n_G-n_x^G)(n_D-n_x^D) .
\end{eqnarray}

\subsection{The noise detector SET}
To illustrate the effect of photon-assisted tunneling clearly we
consider our setup biased so that sequential tunneling is blocked
in one of the SETs (the {\em noise detector}). This is done by
gating $n_x^D < 1/2$ such that the Coulomb energy difference
$\Delta=E_D(1-2n_x^D)$ between the ground state $n_D=0$ and first
excited state $n_D=1$ is larger than temperature and applied
bias $\Delta > \{eV_D^{\text{sd}}/2, k_B T\}$. Neglecting the coupling to
the other SET, the direct current through the detector is given by
cotunneling
\begin{equation}
\label{cotunnelcurrent} I^D_{\text{cot}}=e\frac{eV_D^{\text{sd}}}{h} \frac{1}{24
\pi^2} \left(\frac{R_K}{R_D}\right)^2
\frac{\left(eV_D^{\text{sd}}\right)^2+\left(2\pi
k_BT\right)^2}{\Delta^2},
\end{equation}
where we disregarded cotunneling through the state $n_D=-1$. We
will find that the current induced by photon-assisted tunneling
can be made orders of magnitude larger than $I^D_{\text{cot}}$.

\subsection{The noise generator SET}
The other SET (the {\em noise generator}) is biased such that a
substantial current is produced by sequential tunneling, but for
simplicity we keep the bias low enough so that only two charge
states are involved. To be specific $n_x^G \approx 1/2$, and
$eV_G^{\text{sd}}/2 < 2 E_G$. The current through the noise generator SET
is
\begin{equation}
I_{\text{seq}}^G=e\frac{eV_G^{\text{sd}}}{4h}\frac{R_K}{R_G}\left(1-e^{-eV_G^{\text{sd}}/2k_BT}\right)^{-1}\;
.
\end{equation}
Each electron tunneling in the forward direction will gain the
energy $eV_G^{\text{sd}}/2$ from the applied bias. Generally the electron
will dissipate its excess energy on the island or in the
reservoirs, but in our setup there is also the possibility to give
the energy to a tunnel event in the detector SET.

\subsection{The effect of interaction}
The low-frequency effect of the interaction between the SETs can
be described as an effective gate charge determined by the charge
state of the other SET
\[n_{x,\text{eff}}^G(n_D)=n_{x}^G-\frac{C_c}{C_D^\Sigma}(n_D-n_x^D),\]
and
\[n_{x,\text{eff}}^D(n_G)=n_{x}^D-\frac{C_c}{C_G^\Sigma}(n_G-n_x^G),\] as
discussed in, e.g., Ref.~\onlinecite{SchaeferPhysicaE2003} and \onlinecite{Turek05}. We want
to minimize this effect, in order to clearly show photon-assisted
tunneling. That means choosing the detector bias low enough so
that sequential tunneling is exponentially suppressed for both
effective gate charges $n_{x,\text{eff}}^D(0)$ and $n_{x,\text{eff}}^D(1)\;$.

We will now estimate analytically the detector current due to
photon-assisted sequential tunneling, driven by photons emitted
from the generator with an energy higher than the Coulomb gap of
the detector. We can then compare these analytical estimates with
the numerical results from our algorithmic diagrammatics. We use
$P(E)$-theory,\cite{PofEtheory} considering the noise generator
SET as the environment of the noise detector SET.

\begin{figure}[t]
\centerline{\includegraphics[width=0.65\columnwidth]{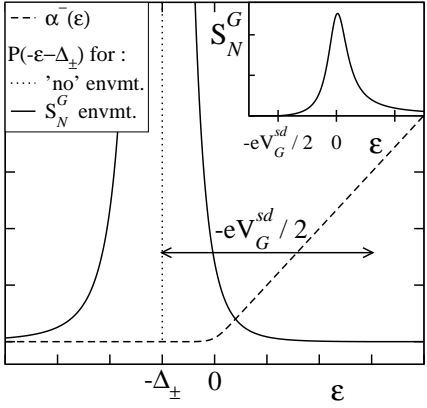}}
\caption{Sketch of functions relevant to calculation of the photon-assisted tunneling rate [Eq.~(\ref{PArateDefEq})].
 Without energy exchange with the environment [$P(\varepsilon)$ is $\delta$ peaked, dotted line] 
the photon-assisted rate is given by the rate function $\alpha^-$ (broken line) at the 
effective energy gap $\Delta_\pm$.
 Coupling to the noise generator SET changes $P(\varepsilon)$ (solid line) and the rate is given by the 
integration in Eq.~(\ref{PArateInt}). The inset shows the asymmetric charge noise of an SET as calculated 
in Ref.~\onlinecite{KackSETquantumNoisePRB2003} 
.}
\label{sketch_P(E)}
\end{figure}

The function
\begin{equation}
P(\varepsilon)=\frac{1}{h}\int_{-\infty}^{\infty} dt
\exp\left(J(t)+i\frac{\varepsilon}{\hbar}t\right) ,
\end{equation}
expresses the probability of exchanging the energy $\varepsilon$
with a certain environment in a single tunnel event. Here the
function
\begin{eqnarray}
J(t)&=&\langle\left[ \hat{\phi}(t)-\hat{\phi}(0) \right]
\hat{\phi}(0)\rangle=\nonumber\\
&=&\frac{E_{\text{int}}^2}{\hbar^2}\int_{-\infty}^{\infty}d\omega\;
\frac{S^G_N(\omega)}{\omega^2} \left(e^{-i\omega
t}-1\right),
\end{eqnarray}
is given by the correlator of the phase fluctuations on the
detector island induced by the electron number fluctuations on the
generator island. 
To simplify the analysis we approximate the fluctuations to be Gaussian.
Then their asymmetric noise spectral density is given by 
\begin{equation}
S^G_N(\omega)=\int_{-\infty}^{\infty} dt\, e^{-i\omega t} \langle
\delta \hat{N}_G(t)\delta \hat{N}_G(0) \rangle .
\end{equation}
We want to calculate the photon-assisted rate for an electron to
tunnel onto the detector island
\begin{eqnarray}
\label{PArateDefEq} \Gamma^{D\pm}_{01}&=&\frac{1}{h}
\frac{R_K}{R_D} \int_{-\infty}^{\infty}  \int_{-\infty}^{\infty}\!\!\!d\varepsilon \, d\varepsilon' \,
f(\varepsilon)\left[1-f(\varepsilon'-\Delta_{\pm})\right]
P(\varepsilon-\varepsilon') \nonumber \\
&=& \frac{1}{h} \frac{R_K}{R_D} \int_{-\infty}^{\infty}
d\varepsilon\; \frac{\varepsilon
P(-\varepsilon-\Delta_{\pm})}{1-e^{-\varepsilon/k_BT}},
\end{eqnarray}
where $\Delta_{\pm}=\Delta\pm eV_D^{\text{sd}}/2$, and $\Delta_\pm$ is the
effective energy gap for an electron tunneling onto the island
from the right/left lead. 
 Thus we need $P(\varepsilon)$ for
$\varepsilon < -\Delta_{\pm}$ indicating the probability to absorb
an energy larger than $\Delta_{\pm}$ from the environment (see Fig.~\ref{sketch_P(E)}). In this
regime, noting that $\Delta_\pm \gg E_{\text{int}}$ and using the short time
expansion $e^{J(t)}=1+J(t)$ we can
approximate\cite{AguadoQuantumNoisePRL2000}
\begin{equation}
P(\varepsilon)=\frac{E_{\text{int}}^2}{h}
\frac{S^G_N(\varepsilon)}{\varepsilon^2} ,\ \ \varepsilon \ll
-E_{\text{int}}.
\label{P(E)_noise}
\end{equation}
In the relevant frequency regime and at zero temperature the
generator SET noise spectrum is
\cite{JohanssonSETquantumNoisePRL2002,DelftProceedings2003,KackSETquantumNoisePRB2003}
\begin{equation}
S^G_{N}(\varepsilon)=\frac{1}{\hbar}
\frac{\Gamma(\varepsilon)}{4\Gamma(0)^2+\varepsilon^2},
\label{SET_noise}
\end{equation}
where
\begin{equation}
\Gamma(\varepsilon)=\frac{1}{2\pi} \frac{R_K}{R_G}
\left[\frac{eV_{G}^{\text{sd}}}{2}+\varepsilon\right] , \ \
|\varepsilon|<\frac{eV_{G}^{\text{sd}}}{2} .
\end{equation}
Furthermore $S^G_{N}(\varepsilon)=0$ for $\varepsilon <
-eV_G^{\text{sd}}/2$, which is related to the fact that one cannot
extract more energy from a single tunnel event than what is given
by the bias voltage.

We thus find the photon-assisted rates for
electrons to tunnel onto the detector island across the right/left
junction (cf.~Fig.~\ref{sketch_P(E)}),
\begin{equation}
\Gamma^{D\pm}_{01} = \frac{1}{2\pi h}\frac{R_K}{R_D}
\int_{0}^{eV_G^{\text{sd}}/2-\Delta_\pm} \!\!\!d\varepsilon \;
\frac{E_{\text{int}}^2}{(\varepsilon+\Delta_\pm)^2}\frac{\varepsilon
\Gamma(-\varepsilon-\Delta_\pm)}
{4\Gamma(0)^2+(\varepsilon+\Delta_\pm)^2}.
\label{PArateInt}
\end{equation}
Here the lower limit of the integral comes from putting 
temperature to zero in Eq.~(\ref{PArateDefEq}) and the upper limit from
the fact that $P(\varepsilon)=0$ for $\varepsilon < -eV_G^{\text{sd}}/2$.
Making the generator bias substantially larger than the detector
gap $(eV_{G}^{\text{sd}} \gg 2\Delta_\pm)$, we find the approximate
photon-assisted rates
\begin{equation}
\Gamma^{D\pm}_{01} = \frac{eV_{G}^{\text{sd}}}{48 \pi^2 h} \frac{R_K}{R_D}
\frac{R_K}{R_G} \frac{E_{\text{int}}^2}{\Delta_\pm^2}
f(2\Gamma(0)/\Delta_\pm),
\label{PATrateresult}
\end{equation}
where
\begin{equation}
f(x)=\frac{3}{x^3}\left(\pi-2\arctan\frac{1}{x}+x\ln{(1+x^2)}-2x\right)
.
\end{equation}
We note that these rates are proportional to the bias voltage of
the noise generator. If the detector gap is not too small we have
$\Gamma(0)\ll \Delta_\pm$ 
which implies $x \ll 1$ and $f(x)$
approaches the limit $f(0)=1$. For simplicity we assume this limit
from here on. The relaxation rates of the detector are\cite{footnote7}
\begin{equation}
\Gamma^{D\pm}_{10} = \frac{R_K}{R_D}\frac{\Delta_\pm}{h},
\end{equation}
and accordingly the probability to find the detector in its
excited state is
\begin{eqnarray}
P_1^D & = &
\frac{\Gamma^{D+}_{01}+\Gamma^{D-}_{01}}{\Gamma^{D+}_{10}+\Gamma^{D-}_{10}}
= \frac{R_K}{R_G} \frac{eV_{G}^{\text{sd}} E_{\text{int}}^2}{48\pi^2 (\Delta_+
+\Delta_-)} \left( \frac{1}{\Delta_+^2}+\frac{1}{\Delta_-^2}
\right) \approx
\nonumber \\
& \approx & \frac{R_K}{R_G} \frac{eV_{G}^{\text{sd}} E_{\text{int}}^2}{48 \pi^2
\Delta^3},
\end{eqnarray}
and the photon-assisted tunneling (PAT) detector current (assuming
$eV_D^{\text{sd}}\ll\Delta_\pm$)
\begin{equation}
I^D_{\text{pa}} = e (P_1^D \Gamma^{D+}_{10}-P_0^D \Gamma^{D+}_{01})
\approx e \frac{R_K}{R_D} \frac{R_K}{R_G} \frac{eV_{G}^{\text{sd}}
eV_{D}^{\text{sd}} E_{\text{int}}^2}{32 \pi^2 h \Delta^3} .
\label{PATresult}
\end{equation}
Comparing this with the standard cotunneling current in the detector,
given in Eqn.~(\ref{cotunnelcurrent}),
\begin{equation}
\frac{I^D_{\text{pa}}}{I^D_{\text{cot}}}=\frac{3}{4} \frac{R_D}{R_G}
\frac{eV_{G}^{\text{sd}}}{\Delta}
\left(\frac{E_{\text{int}}}{eV_{D}^{\text{sd}}}\right)^2 ,
\end{equation}
we find that the photon-assisted detector current can be made
substantially larger than the usual cotunneling current by using
high resistance tunnel junctions in the detector and a weak
detector bias. 

\subsection{Results of diagrammatic technique}
With the diagrammatic technique developed above, we have a method at our disposal for analyzing 
the complete two-island system on an equal footing. We do not perform a separation
into detector and noisy environment as with the $P(E)$-theory.
In particular, we can study the mutual influence of the two SETs also for the strong coupling case.

In the following, we will nonetheless focus on the weak coupling case; here we can clearly show the 
effects of PAT and sensibly compare to the results of $P(E)$-theory [Eq.~(\ref{PATresult})]. Low-frequency effects, as
described by the effective gate charges, and back-action are all intrinsically included in the calculation, but parameters 
are chosen to minimize these.

\begin{figure}[h]
\centerline{\includegraphics[width=0.98\columnwidth]{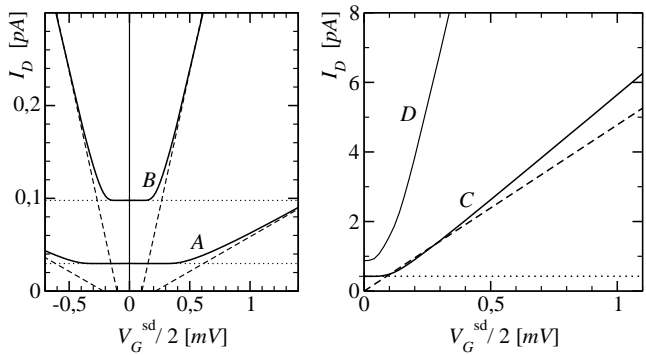}}
\caption{Results of the diagrammatic technique for the generator/detector setup (solid lines) 
compared to standard cotunneling (dotted lines) and photon-assisted current with slope as predicted from
$P(E)$ theory (dashed lines, x-intercepts fitted). All curves at $T=25\,\rm mK$, $eV^{\text{sd}}_D = 0.5\,k_B\rm K$, and $n_{x,\text{eff}}^G(n_D=0)=0.5$ with $\alpha^j_0=0.01$ for each junction;
 $E_G= 10\,k_B\rm K$, $E_D= 10\,k_B\rm K$ for $A\,$--$\,C$ and $5\,k_B\rm K$ for $D$, $E_{\text{int}}= 0.5\,k_B\rm K$ for $A\,$--$\,C$ and $1\,k_B\rm K$ for $D$. $n_x^D$ as 
follows: $A$ --- 0.3, $B$ --- 0.4, $C$ --- 0.45, $D$ --- 0.4.
}
 \label{noise_det}
\end{figure}

Figure~\ref{noise_det} (solid lines) shows current in the detector SET upon changing the bias on the generator SET for four different 
sets of sample parameters and gating. The detector is gated close to degeneracy point, i.e., 
$n_{x,\text{eff}}^G(n_D=0)=0.5$, as the detector is set preferentially unoccupied in the Coulomb blockade region.
The detector is to remain in this blockade regime, independent of the generator's charge state.
To achieve considerable, experimentally detectable, current through the detector, we must not be ``too deep'' in Coulomb blockade.
Hence, we move from a region of well-defined blockade in curve $A$ of Fig.~\ref{noise_det} to a ``mixed'' regime in curve $D$ ($E_{\text{int}}\approx \Delta_\pm$) by
 changing gating (and coupling). The exact parameters are given in the captions of Fig.~\ref{noise_det}.

For \emph{small driving} of the generator, we find the standard cotunneling current in the detector SET. The cotunneling 
current shown in Fig.~\ref{noise_det} (dotted horizontal lines for curves 
$A$--$C$) is calculated for a single SET with effective gating 
$n_{x,\text{eff}}^D(n_G)$, weighted with the probabilities for the two generator states $P(n_G=0)\approx 1/2 \approx P(n_G=1)$.

For \emph{strong driving} the condition for PAT above is fulfilled (i.e., $eV_G^{\text{sd}} \gg 2\Delta_\pm$) and we find the 
detector current increasing proportional to the driving voltage of the 
generator [cf.~Eq.~(\ref{PATrateresult})]. For the curves $A$ and $B$ we find 
the slope as predicted by $P(E)$-theory [Eq.~(\ref{PATresult}), dashed lines in $A$--$C$]. For curves $C$ and $D$ we do not meet all assumptions made in the 
derivation of the PAT rates. Consequently PAT and low-frequency interaction effects intermingle in a complex manner and the result
deviates from Eq.~(\ref{PATresult}). For $D$ the coupling is chosen so strong, that back-action effects are also  relevant, i.e., a meaningful 
definition of effective energy gaps and, therefore, comparison to PAT and standard cotunneling results is not possible anymore.   

As the photon-assisted rate was calculated assuming strong generator driving only, the crossover between standard cotunneling and PAT is not properly described.
The results of our full theory (solid lines in Fig.~\ref{noise_det}) show, however, the expected behavior\---an onset 
of PAT (i.e., the x-intercept of the dashed lines
in Fig.~\ref{noise_det} curves $A$ and $B$), where $eV_G^{\text{sd}}\approx\Delta_\pm$ and a corresponding crossover, when the PAT rate reaches the standard cotunneling rate.

In conclusion our theory for the generator/detector setup shows standard cotunneling for low generator bias and photon-assisted tunneling for high generator bias, where the crossover depends on the effective energy gap of the detector. 
For weak generator-detector coupling our results agree perfectly with a $P(E)$ calculation of photon-assisted sequential tunneling, treating the generator as an energy providing environment of the detector. While the relative effect is strong\---PAT current is of the same order as the standard cotunneling current\---the overall currents are quite small. Stronger coupling yields higher currents, but now current is not given by photon-assisted sequential tunneling alone and our simplified $P(E)$ calculation is insufficient.

In the example of a detector-generator setup above we choose parameters to a regime where we can isolate one particular new type of cotunneling process (involving energy exchange between islands) as dominant constituent of current. Then we could compare to a $P(E)$ calculation of photon-assisted sequential tunneling, simple enough to achieve analytical and physical transparent results. To reach such simple analytical results we made use of a number of assumptions and approximations:
Within the $P(E)$ calculation we did not consider standard cotunneling through the detector island (nor through the generator island). 
In calculating the noise spectrum of the generator determining $P(E)$ we neglect its coupling to the detector, which is permissible for appropriately chosen parameters. The Gaussian approximation was used connecting the autocorrelation of phase fluctuations on the detector island to the asymmetric charge noise of the generator, while higher moments were neglected. The influence of higher moments (and consequently their detection) has been studied in a number of other systems recently.\cite{NazarovQuantumNoiseDetectionPRL2004, SoninNoiseDetectionPRB2004,
PekolaNoiseDetectionPRL2004, HakonenNoiseDetectionPRL2004,
HeikkilaQuantumNoiseDetectionPRL2004,GrabertPRL05}
To achieve analytically tractable results we exercised a short time expansion, which is justified in the weak coupling regime. 
Finally, the zero temperature and large bias limit was assumed.  

The latter two approximations are in no way crucial and for convenience and simplicity of results only. Reaching such transparent results for comparison to the perturbative approach is the main purpose of the $P(E)$ analysis we performed here. Relinquishing this notion of achieving simple analytical results, a $P(E)$ analysis may also be further refined to loosen some of the restrictions imposed above: 
Consider cotunneling through the detector island, which depends on the state of the generator island (by the effective gating), which is occupied with either zero or one electron with some probability. This results in a shift of the Coulomb gap of the detector island dependent on the generator state. 
A $P(E)$-type theory of cotunneling in the detector can be set up to incorporate that  effect of the environment.
This will then just correspond to the  ``standard cotunneling'' results in Fig.~\ref{noise_det}, which takes this very effect into account.
In principle a $P(E)$ analysis can also be employed without making use of the short time expansion to extend towards strong coupling. However, such an approach is limited by the increasing importance of backaction effects in the strong coupling regime. 
Indeed, by construction backaction effects are not included in a $P(E)$-theory. Its basic principle is to describe the environment by a single function [namely $P(E)$], which in consequence cannot depend on the system's state dynamics. Backaction effects are, however, fully accounted for within the perturbative approach, where there is no separation into system and environment parts as in $P(E)$. 

It should be noted that we implemented a perturbative approach up to second order in the coupling. In the scenario above, the main contribution to current stems from sequential tunneling in the detector (proportional to the detector coupling) assisted by fluctuations due to sequential tunneling through the generator (proportional to the generator coupling). By this way of counting powers of the coupling, it is obvious, that a second-order analysis cannot capture the influence on detector current of higher-order quantum fluctuations in the generator.

We have demonstrated that cotunneling contributions involving tunneling in different islands, where energy is exchanged between electrostatically coupled SETs, can significantly contribute to transport. A normal state SET is therefore inherently sensitive to finite frequency noise of charge, to which it couples electrostatically. A number of other systems have recently been suggested and used as
on-chip measurement devices for finite frequency current and/or
voltage noise: a single\cite{HartmannNazarov} or double quantum 
dot,\cite{AguadoQuantumNoisePRL2000} a superconducting 
qubit,\cite{ClerkQubitSpectrometer2003} or a tunnel junction, both in the
normal and superconducting state.\cite{DeblockQuantumNoiseDetectionScience2003,
NazarovQuantumNoiseDetectionPRL2004, SoninNoiseDetectionPRB2004,
PekolaNoiseDetectionPRL2004, HakonenNoiseDetectionPRL2004,
HeikkilaQuantumNoiseDetectionPRL2004,GrabertPRL05} A normal state SET is inherently sensitive to integrated noise of a broad energy window, due to the peculiarities of the sequential tunneling rates [see Eqs.~(\ref{PArateDefEq}) and (\ref{PArateInt})]. In contrast, other systems mentioned above are particularly designed to make use of sharp resonancelike features in the detector to have high sensitivity for noise detection of a certain frequency.

\section{Summary and conclusions}
In this paper we extended the real-time diagrammatics of Ref.~\onlinecite{SchoellerSchon94} to model transport 
through metallic multi-island systems. We discussed our approach to automatically generate and compute
diagrams in executing a systematic perturbation expansion up to second order in the tunneling conductance. 
This corresponds to sequential and cotunneling terms. Convergence properties of the perturbative expansion were analyzed.

In a setup of two coupled SETs, we demonstrated the importance of cotunneling involving both islands, where energy can be exchanged in between the two tunneling processes. This is linked to the notion of photon-assisted tunneling. We performed a $P(E)$ analysis, treating one SET as an environment for the other one and found excellent agreement between both approaches in the relevant weak coupling limit. In this regime one SET works as a detector of the finite frequency charge noise of the other SET.
  
\begin{acknowledgments}
We acknowledge helpful discussions with R.~Sch\"afer and financial support by DFG via Graduiertenkolleg 726.
\end{acknowledgments}

\appendix
\section{Calculating diagrams\label{calculatingdiagrams}}
In this appendix we explain how integral expressions,
corresponding to the diagrams created, are efficiently calculated.
First of all, the number of expressions we have to evaluate can be
reduced by identifying several distinct types, only differing in
some parameters. For these small number of terms numerical and
analytical methods of automated calculation are discussed.

\subsection{Reducing denominators}
\begin{figure}[]
\centerline{\includegraphics[width=0.7\columnwidth]{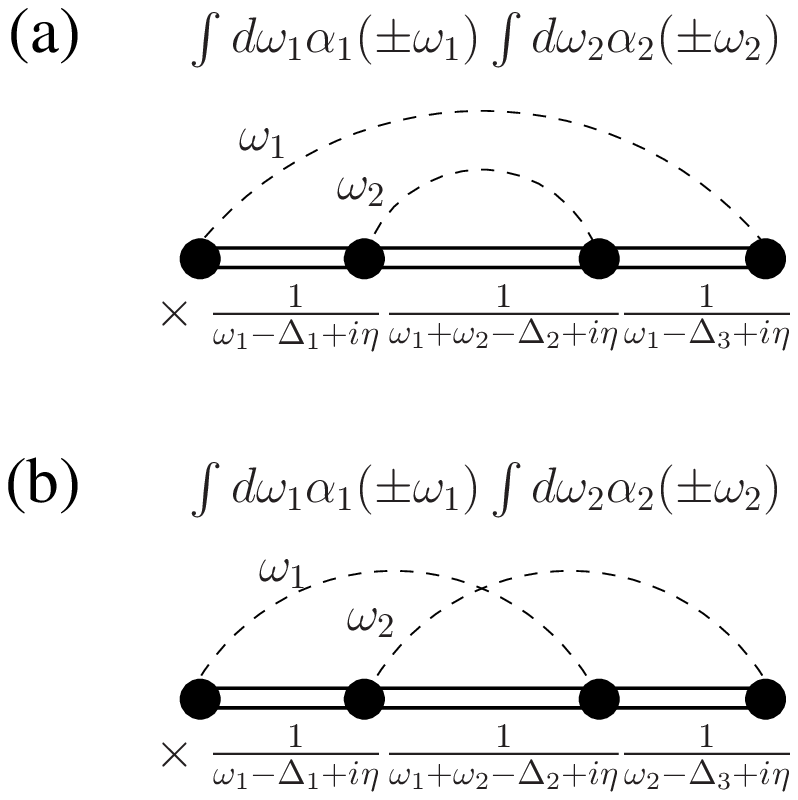}}
\caption{Two topologically distinct types of diagrams are found.
This yields four different integral terms, to be evaluated.}
\label{denominators}
\end{figure}

The general structure of integral expressions for a diagram follows easily 
from the diagrammatic rules discussed above. A self-energy diagram
of order $R$ features $R$ integrals over contraction line energies
$\omega_r$ with corresponding rate functions
$\alpha^{j\pm}(\omega_r)$ but also $(2R-1)$ denominator factors
$[\Delta E +i \eta]^{-1}$ for each segment in between two
vertices. The workhorse for evaluating these expressions is
Cauchy's formula,
\begin{equation}
\frac{1}{x+i\eta}=P\frac{1}{x}-i\pi\delta(x) ,
\label{Cauchy1}
\end{equation}
which obviously can only be used when integrating over the
variable $x$. Consequently we have to reduce the number of
denominator terms from $(2R-1)$ to $R$. Note here, that Cauchy's
formula can still be applied to higher-order poles,
\begin{equation}
\frac{1}{(x+i\eta)^2}=-\frac{d}{dx}\;\frac{1}{x+i\eta}= -\frac{d}{dx}\left[ P\frac{1}{x}-i\pi\delta(x) \right] .
\label{Cauchy2}
\end{equation}
While this reduction is possible for arbitrary order, we will
illustrate it for second-order diagrams only. 
As illustrated in  Fig.~\ref{denominators}, there are two topologically distinct diagrams, depending 
on whether the first vertex is connected to third or fourth vertex. Note, that the 
position of vertices on the upper or lower branch is irrelevant for the general structure of the analytical 
expressions, which are indicated  in Fig.~\ref{denominators}. 
Elementary algebraic manipulation of the denominator factors\---merely the
treatment of infinitesimal $\eta$ requires some care\---yields four expressions 
with the number of denominator factors reduced as
required. 

From the expression in Fig.~\ref{denominators}(a) we get
\begin{eqnarray*}
&& \frac{1}{\omega_1-\Delta_1+i\eta}\,\frac{1}{\omega_1+\omega_2-\Delta_2+i\eta}\,\frac{1}{\omega_1-\Delta_3+i\eta} \Bigg.\\
&&= \left\{ \begin{array}{r@{\quad:\quad}l}  
 \Delta_1 = \Delta_3 & \displaystyle \frac{d}{d\Delta_1} \frac{1}{\omega_1-\Delta_1+i\eta}\, \frac{1}{\omega_1+\omega_2-\Delta_2+i\eta} \bigg.\\
 \Delta_1\neq \Delta_3 &  \begin{array}[t]{l} \displaystyle \frac{1}{\Delta_1-\Delta_3}\,\frac{1}{\omega_1+\omega_2-\Delta_2+i\eta}  \Bigg.\\
\times \displaystyle \left(\frac{1}{\omega_1-\Delta_1+i\eta}    - \frac{1}{\omega_1-\Delta_3+i\eta}  \right)\bigg.
\end{array}
\end{array}
\right. .
\end{eqnarray*}
Correspondingly, Fig.~\ref{denominators}(b) yields
\begin{eqnarray*}
&& \frac{1}{\omega_1-\Delta_1+i\eta}\,\frac{1}{\omega_1+\omega_2-\Delta_2+i\eta}\,\frac{1}{\omega_2-\Delta_3+i\eta} \Bigg.\\
&&\hspace*{-0.3cm}= \left\{ \begin{array}{r@{\;:\;}l}  
 \Delta_1 + \Delta_3 = \Delta_2 & \begin{array}[t]{l} \displaystyle \frac{d}{d\Delta_2} \frac{1}{\omega_1 + \omega_2 -\Delta_2+i\eta} \Bigg.\\
\times \displaystyle \left(\frac{1}{\omega_1-\Delta_1+i\eta} + \frac{1}{\omega_2-\Delta_3+i\eta} \right)\Bigg.
\end{array} \\
\Delta_1 + \Delta_3 \neq \Delta_2 & \begin{array}[t]{l} 
\displaystyle \left(\frac{1}{\omega_1 + \omega_2 -\Delta_1 - \Delta_3 +i\eta} \right.  \Bigg.\\
\displaystyle \;\;\;\left.  - \frac{1}{\omega_1 + \omega_2 -\Delta_2+i\eta}  \right) \frac{1}{\Delta_1 + \Delta_3 - \Delta_2 } \bigg. \\
\displaystyle \times  \left(\frac{1}{\omega_1-\Delta_1+i\eta}    +  \frac{1}{\omega_2-\Delta_3+i\eta}  \right) \Bigg.\\
\end{array}
\end{array}
\right. 
\end{eqnarray*}
for the second topologically distinct type.

\subsection{Mirror rule}
Further simplifications arise from an inherent symmetry property
of the diagrammatic rules. From the construction principles for
diagrams it is easily seen that, starting from any diagram, we
will obtain another possible diagram by the following operation:

Reflect the diagram along a horizontal line, whereby exchanging
the forward and backward branches of the contour, and change the
direction of all tunneling lines.
 The charge state for any part of
the contour then remains unchanged. Both diagrams contribute to
the same matrix entry of the self-energy, as only the diagonal
part of $\Sigma$ has to be considered.
Application of the diagrammatic rules shows that merely the
energies in the denominator terms change sign,
\begin{equation}
\frac{1}{\Delta E+i\eta} \rightarrow  \frac{1}{-\Delta E+i\eta} = - \left( \frac{1}{\Delta E+i\eta} \right)^* ,
\end{equation}
resulting in cancellation of the real parts of the two mirrored
diagrams.
 Consequently we only calculate the imaginary part of
each diagram. This allows to utilize the $\delta$-function part of  Cauchy's formula 
for evaluating one of the two integrations. Our problem reduces then to the calculation of a
single (principal value) integral, where the resulting integrand can be written as a product of one or two
rate functions\---there might be a derivative acting on one of
them\---multiplied with the principal value term
$P\frac{1}{\omega}$.

\subsection{Evaluating the integrals\label{evaluatingintegrals}}
Above [see Eq.~(\ref{rate_definition})] we calculated analytical expressions for
the rate functions for the case of normal reservoirs with a
constant density of states (DOS) and equal temperatures for both sides
of the tunneling barrier. It is interesting, though, that
important general features of the rate functions will
persist, if these conditions are relaxed, e.g., for
nonequal temperatures, relevant for considering thermopower or
self-heating effects, or for tunneling rates of quasiparticles in
superconducting devices, where the gapped BCS density of states
enters. Common to all these cases, however, is the existence of a
certain onset energy, where the asymptotic behavior is linear on
one side, and vanishing on the other side of this threshold; the asymptotic
convergence is hereby governed by some Boltzmann factor.

Evidently, a high-energy cutoff is needed for convergence of
integrals, where the two rate functions grow in the same
direction. This cutoff is provided for in a natural way by the
bandwidth of the reservoirs, as a more careful analysis of the
rate function for a DOS limited to some finite bandwidth reveals. The
cutoff will at most enter logarithmically into any final result\---indeed, for the single SET, it is known to drop out completely
from the measurable quantities current and average charge. Hence,
there is no need for a microscopically detailed derivation of a
specific cutoff function and we can choose for convenience of
calculation a Breit-Wigner factor centered around the onset.

An \emph{analytical} solution of the integrals is then possible
for the rate functions as given in Eq.~(\ref{rate_definition}). Contour integration
leads to sums over Matsubara frequencies, which in turn results in
analytical expressions involving Digamma functions, where the
onset energies enter as parameters (cf.~Ref.~\onlinecite{KoenigSchoellerSchoen97}).

Since further algebraic manipulation of these complex terms is not
feasible for multi-island geometries, it can be more convenient to
adopt a \emph{numerical} approach, already for calculating the
integrals\---the additional benefit being that this approach is
also capable of effortlessly dealing with the alternative rate functions
discussed above. Numerical evaluation is helped by
the general features shared by all these rate functions. They
allow for precise numerical integration for a small region (of a
few $k_BT$) around the onset, while the asymptotics can be
trivially integrated analytically.

\section{Convergence of a perturbative expansion\label{convergence}}
In this appendix we address the question of the applicability range of 
our perturbation expansion.
To motivate a criterion for general multi-island systems let us first
consider the case of a single SET.
In this case, we have a \emph{nonperturbative} result for the current within 
the so-called resonant tunneling approximation (RTA)\cite{SchoellerSchon94} 
at hand,
\begin{equation}
I^{\text{sd}} =\frac{2\pi e}{\hbar} \int d\omega \;\alpha_L(\omega)\alpha_R(\omega) \left|\Pi(\omega)\right|^2 \left[ f_L^+(\omega)- f_R^+(\omega) \right] ,
\label{current_RTA1}
\end{equation}
where $\alpha_r=\alpha_r^+ + \alpha_r^-$ and the propagator $\Pi(\omega) = \left[ \omega - \Delta -\sigma(\omega) \right]^{-1}$
 dressed with first-order transitions as exemplified in the self-energy
\begin{eqnarray}
\text{Re}\, \sigma(\omega) &=& -2\sum_r\alpha_0^r(\omega-\mu_r)\\
 & & \quad \times \left[\ln\left(\frac{\beta E_C}{2\pi}\right)-\text{Re}\,\Psi\left(\frac{i\beta(\omega-\mu_r)}{2\pi}\right)\right],\nonumber\\
\text{Im}\, \sigma(\omega) &=& -\pi\sum_r\alpha^r(\omega) \bigg. ,
\end{eqnarray}
where the cutoff energy $E_C$ is of the order of the charging energy.
We see that the dominant contribution to the current comes from energies
around the charging energy $\Delta$.
Higher-order quantum fluctuations embodied by $\sigma(\omega)$ both shift
(real part) and broaden (imaginary part) the region of contributing energies.

For definiteness we consider now the scenario $\mu_L - \mu_R =eV >0$, where we are interested in the onset of 
sequential tunneling around 
$\mu_L \lesssim \Delta$, while a finite 
bias $eV \gg k_BT$ is applied asymmetrically across the SET. The main contribution to the current in this 
regime is then given by electrons slowly tunneling onto the island from the left lead (the rate for this 
bottleneck process being $\alpha_L^+(\omega)$, while they quickly tunnel off to the right lead [$\alpha_R^-(\omega) \gg \alpha_L^+(\omega)$].
Consequently the current [Eq.~(\ref{current_RTA1})] simplifies to
\begin{equation}
I^{\text{sd}} =-\frac{2e}{\hbar}\, \text{Im}\, \int d\omega \; \frac{\alpha_L^+(\omega)}{\omega - \Delta -\sigma(\omega)}  ,
\label{current_RTA2}
\end{equation}
where the imaginary part of the self-energy, the lifetime width of the resonant level, is dominated by 
the (decay) rate for tunneling out to the right lead.

To understand the effect of a systematic perturbation expansion of the current operator, as performed in the main part of the paper, 
we expand now this expression for the current [Eq.~(\ref{current_RTA2})] in orders of the tunneling conductance $\alpha_0$.
Expansion of the denominator yields
\begin{eqnarray}\label{expansion}
\frac{1}{\omega - \Delta -\sigma(\omega)} \!\!&=&  \!\! \frac{1}{\omega - \Delta + i \eta} \\
&& \!\!\!\!  +  \frac{\sigma(\omega)}{(\omega - \Delta +i\eta)^2} +  
\frac{1}{2} \frac{\sigma(\omega)^2}{(\omega - \Delta+i\eta)^3} + \hdots \;, \nonumber
\end{eqnarray}
where the denominator terms are regularized by infinitesimal imaginary parts.
This should be construed in terms of the Cauchy identity for generalized 
functions 
and derivatives thereof [Eqs.~(\ref{Cauchy1}) and (\ref{Cauchy2})].

\begin{figure}[]
\centerline{\includegraphics[width=0.8\columnwidth]{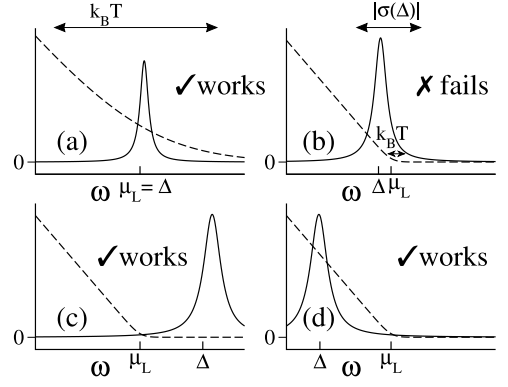}}
\caption{Sketch of $\alpha_L^+(\omega)$ (broken line) and  $\text{Im}\,\Pi(\omega)$ (solid line) in the integrand of Eq.~(\ref{current_RTA2}). 
Perturbative expansion works, if the rate $\alpha_L^+$ is approximately linear in  the region given by the peak in $\text{Im}\,\Pi(\omega)$, corresponding to $\left|\sigma(\Delta)\right| < k_BT$ on resonance (a) or to  $\left| \Delta -\mu_L \right| > \left| \sigma(\Delta)\right|$ in (c) and (d). The expansion fails (b) for  $\text{Im}\,\Pi(\omega)$ incorporating the kink in $\alpha_L^+$.
} \label{sketch_convergence}
\end{figure}

The term of this expansion of order $\alpha_0^n$ contains the denominator $(\omega -\Delta +i\eta)^{-(n+1)}$, which results in a $n^{\text{th}}$ derivative of a $\delta$--function. Within the integral expression for the current [Eq.~(\ref{current_RTA2})] we will use partial integration, so that the derivatives act on $\alpha_L^+(\omega)\sigma(\omega)^{n}$. The expansion in terms of $\alpha_0$ therefore correspond to a Taylor expansion around the bare resonance $\Delta$. In Fig.~\ref{sketch_convergence} we visualize the integration of Eq.~(\ref{current_RTA2}) for different parameter regimes. We integrate the rate function  $\alpha_L^+(\omega)$, which has a kink at $\omega = \mu_L$ with a characteristic width of $k_BT$, multiplied with $\text{Im}\,\Pi(\omega)$, which is peaked close to $\Delta$, shifted by $\left|\text{Re}\,\sigma(\omega)\right|$ and has a width given by $\left|\text{Im}\,\sigma(\omega)\right|$. 

As just shown, the perturbation expansion consists of expanding the peak of  $\text{Im}\,\Pi(\omega)$ in terms of derivatives of $\delta$--functions, consequently capturing properies of $\alpha_L^+(\omega)$ (and its derivatives) at $\Delta$ only.

As is intuitively clear from Figs.~\ref{sketch_convergence}(a)--\ref{sketch_convergence}(d), this will work, if the peak $\text{Im}\,\Pi(\omega)$ is sharp, as compared to the structure in  $\alpha_L^+(\omega)$ in the relevant regime.
This means that either $\left|\sigma(\Delta)\right|$ has to be smaller than $k_BT$ for considering the system on resonance $\Delta=\mu_L$ [Fig.~\ref{sketch_convergence}(a)] or that the expansion takes place so far away from resonance with the left lead, that the peak $\text{Im}\,\Pi(\omega)$ is not incorporating the kinked region of $\alpha_L^+(\omega)$; i.e., $\left|\Delta -\mu_L \right| > \left| \sigma(\Delta)\right|$ [see Fig.~\ref{sketch_convergence}(c) and \ref{sketch_convergence}(d)]. Then the rate function is expanded in the asymptotic region, where it is exponentially suppressed [Fig.~\ref{sketch_convergence}(c)] or rising approximately linearly [\ref{sketch_convergence}(d)]. Consequently higher-order derivatives from higher-order terms of the expansion in $\alpha_0$ do not contribute and second-order perturbation theory suffices.

We can formalize these intuitive arguments by considering the Taylor expansion of the bottleneck rate, resulting from the derivatives of the $\delta$--function in the perturbative expansion [Eq.~(\ref{expansion})],
\begin{equation*}
\frac{\alpha_0^L \omega}{\exp{(\beta \omega)} - 1} = \alpha_0^L \left( \frac{1}{\beta} + \frac{\omega}{2} + \frac{\beta\omega^2}{6 \cdot 2!} - \frac{\beta^3\omega^4}{30 \cdot 4!} + \hdots \right). 
\end{equation*} 
Note the increasing powers of the inverse temperature $\beta$, endangering convergence for low temperatures.
The largest term now to appear in the $n^{\text{th}}$ term of the $\alpha_0$ expansion is a term  with magnitude
\begin{equation*}
\partial^{n-1}_{\omega} \alpha_L^+(\omega) \sigma(\omega)^{n-1} \propto \alpha_0^L \beta^{n-2}\left[\sigma(\Delta)\right]^{n-1} ,
\end{equation*}  
where all derivatives have acted on $\alpha_L^+(\omega)$.\cite{footnote8} 
Correspondingly we get a series in the parameter $\beta\left|\sigma(\Delta)\right|$, which has to be small for convergence, confirming our argument above.

In summary, we found a criterion for the applicability range of second-order 
perturbation theory.
The ideas developed by considering a single SET can be easily generalized to
an arbitrary multi-island geometry.
For each island, all tunneling rates should be approximately linear in the
region of the contributing energies.
This leads to the condition
\begin{equation} \label{criterion}
\max\{k_BT, \left|\Delta-\mu_j\right|\} > \left|\sigma(\Delta)\right| ,
\end{equation}
where $\sigma(\Delta)$ is the self energy characterizing shift and width of
the resonance due to quantum fluctuations.
For $k_BT > \sigma(\Delta)$  the tunneling rate is approximately linear on the relevant scale of integration, while for
 $\left|\Delta - \mu_j\right| > \sigma(\Delta)$ the kink in the rate function is outside the integration region, and 
consequently the rate is either linear or exponentially small.

An example of the failure of perturbative expansion is an SET at finite bias 
 and low temperature. Likewise in a setup of two islands in series\cite{SchaeferSeriesPRB2005} the resonance is shifted by a self-energy scaling as $\text{Re}\,\sigma \propto \alpha_0 E_C$, not vanishing for aligned levels of the two islands. This can result in a failure of perturbative expansion at low temperatures.\cite{TeemuPRB99}

No problems are encountered for the scenario laid out in Sec.~\ref{example}. Here the generator SET is completely dominated by sequential tunneling, while the detector, biased on the order of temperature, correspondingly yields a small $\left|\sigma(\Delta)\right| \approx \alpha_0 k_BT$, fulfilling Eq.~(\ref{criterion}).


\begin{thebibliography}{99}

\bibitem{FultonDolan87}
T.~A.~Fulton and G.~J.~Dolan, Phys.~Rev.~Lett.~{\bf 59}, 109 (1987).

\bibitem{KulikShekhter75}
O.~Kulik and R.I.~Shekhter, Sov.~Phys.~JETP {\bf 41}, 308 (1975).

\bibitem{AverinLikharev86}
D.V.~Averin and K.~K.~Likharev, J.~Low Temp.~Phys. {\bf 62}, 345 (1986).

\bibitem{GrabertDevoretBook92}
{\it Single Charge Tunneling}, edited by H.~Grabert and M.~H.~Devoret, NATO Advanced Study Institute Series B: Physics (Plenum Press, New York, 1992), Vol.~{\bf 294}.

\bibitem{Amman89}
M.~Amman, E.~Ben-Jacob, and K.~Mullen, Phys.~Lett.~A {\bf 142}, 431 (1989).

\bibitem{AverinKorotkovNazarov91}
D.~V.~Averin, A.~N.~Korotkov, and Yu.~V.~Nazarov, Phys.~Rev.~Lett.~{\bf 66}, 2818 (1991).

\bibitem{Korotkov94}
A.~N.~Korotkov, Phys.~Rev.~B {\bf 50}, 17674 (1994).

\bibitem{TeemuPRB99}
T.~Pohjola, J.~K\"onig, H.~Schoeller, and G.~Sch\"on, Phys.~Rev.~B
{\bf 59}, 7579 (1999).

\bibitem{DesteveBook92}
D.~Esteve, in {\it Single Charge Tunneling} (Ref.~\onlinecite{GrabertDevoretBook92}), p.~109.

\bibitem{SchaeferPhysicaE2003}
R.~Sch\"afer, B.~Limbach, P.~vom~Stein, C.~Wallisser, Physica E (Amsterdam) {\bf
18}, 87 (2003).

\bibitem{Lehnert03}
K.~W.~Lehnert, B.~A.~Turek, K.~Bladh, L.~F.~Spietz, D.~Gunnarsson, P.~Delsing, and R.~J.~Schoelkopf, 
Phys.~Rev.~Lett.~{\bf 91}, 106801 (2003).

\bibitem{Turek05}
B.~A.~Turek, K.~W.~Lehnert, A.~Clerk, D.~Gunnarsson, K.~Bladh, P.~Delsing, and R.~J.~Schoelkopf,
 Phys.~Rev.~B {\bf 71}, 193304 (2005).

\bibitem{DelsingLikharev89}
L.~S.~Kuzmin, P.~Delsing, T.~Claeson, and K.~K.~Likharev, Phys.~Rev.~Lett.~{\bf 62}, 2539 (1989).

\bibitem{SaclayFirstPumpPhysicaB91}
H.~Pothier, P.~Lafarge, P.~F.~Orfila, C.~Urbina, D.~Esteve, and M.
H.~Devoret, Physica B {\bf 169}, 573 (1991).

\bibitem{PumpAndTurnstileIEEE91}
C.~Urbina, H.~Pothier, P.~Lafarge, P.~F.~Orfila, D.~Esteve, M.
Devoret, L.~J.~Geerligs, V.~F.~Anderegg, P.~A.~M.~Holweg, and J.
E.~Mooij, IEEE Trans.~Magn.~{\bf 27}, 2578 (1991).

\bibitem{Pekola_thermometry94}
J.~P.~Pekola, K.~P.~Hirvi, J.~P.~Kauppinen, and M.~A.~Paalanen, Phys.~Rev.~Lett.~{\bf 73}, 2903 (1994).

\bibitem{Martinis7PumpErrorsPRB1999}
R.~L.~Kautz, M.~W.~Keller, and J.~M.~Martinis, Phys.~Rev.~B {\bf
60}, 8199 (1999).

\bibitem{KellerZimmermanStandardReview2003}
N.~M.~Zimmerman and M.~W.~Keller, Meas.~Sci.~Technol.~{\bf 14},
1237 (2003).

\bibitem{PTB04}
H.~Scherer, S.~V.~Lotkhov, G.-D.~Willenberg, and A.~B.~Zorin, IEEE Trans.~Instrum.~Meas.~{\bf 54}, 666 (2005).

\bibitem{DelsingNature05}
J.~Bylander, T.~Duty, and P.~Delsing, Nature {\bf 434}, 361 (2005).

\bibitem{SchaeferSeriesPRB2005}
B.~Limbach, P.~vom~Stein, C.~Wallisser, and R.~Sch\"afer, 
Phys.~Rev.~B {\bf 72}, 045319 (2005).

\bibitem{Likharev87}
K.~K.~Likharev, IEEE Trans.~Magn.~{\bf 23}, 1142 (1987).

\bibitem{AverinOdintsovPhysLettA1989}
D.~V.~Averin and A.~A.~Odintsov, Phys.~Lett.~A {\bf 140}, 251 (1989).

\bibitem{Mqt-tunneling}
D.~V.~Averin and Yu.~V.~Nazarov, Phys.~Rev.~Lett. {\bf 65}, 2446 (1990).

\bibitem{JensenMartinisHighOrderApproxPRB1992}
H.~D.~Jensen and J.~M.~Martinis, Phys.~Rev.~B {\bf 46},
13407 (1992).

\bibitem{Matveev91}
K.~A.~Matveev, Sov.~Phys.~JETP {\bf 72}, 892 (1991),
[Zh.~Eksp.~Teor.~Fiz.~{\bf 99}, 1598 (1991)].

\bibitem{Joyez97}
P.~Joyez, V.~Bouchiat, D.~Esteve, C.~Urbina, and M.~H.~Devoret, Phys.~Rev.~Lett.~{\bf 79}, 1349 (1997).

\bibitem{Wallisser02}
C.~Wallisser, B.~Limbach, P.~vom Stein, R.~Sch\"afer, C.~Theis, G.~G\"oppert, 
and H.~Grabert, Phys.~Rev.~B {\bf 66}, 125314 (2002).

\bibitem{KoenigSchoellerSchoen97}
J.~K\"onig, H.~Schoeller, and G.~Sch\"on, Phys.~Rev.~Lett.~{\bf 78}, 4482 
(1997); Phys.~Rev.~B~{\bf 58}, 7882 (1998).

\bibitem{GolubevZaikin96}
D.S.~Golubev and A.D.~Zaikin, Pis'ma Zh.~Eksp.~Teor.~Fiz.~{\bf 63}, 953 (1996)
[JETP Lett.~{\bf 63}, 1007 (1996)].

\bibitem{GoeppertGrabert98}
G.~G\"oppert and H.~Grabert, Phys.~Rev.~B {\bf 58}, R10155 (1998).

\bibitem{KoenigSchoeller98}
J.~K\"onig and H.~Schoeller, Phys.~Rev.~Lett.~{\bf 81}, 3511 (1998).

\bibitem{GoeppertHuepperGrabert00}
G.~G\"oppert, B.~H\"upper, and H.~Grabert, Phys.~Rev.~B~{\bf 62}, 9955 (2000).

\bibitem{SchoellerSchon94}
H.~Schoeller and G.~Sch\"on, Phys.~Rev.~B {\bf 50}, 18436 (1994);
J.~K\"onig, H.~Schoeller, and G.~Sch\"on, Europhys.~Lett.~{\bf 31}, 31 (1995).

\bibitem{multi-island-numerics-Fonseca-1995}
L.~R.~C.~Fonseca, A.~N.~Korotkov, K.~K.~Likharev, and A.~A.
Odintsov, J.~Appl.~Phys.~{\bf 78}, 3238 (1995).

\bibitem{Simon_by_ChristophWasshuber}
C.~Wasshuber, {\it Computational Single-Electronics} (Springer-Verlag, Berlin, 2001).

\bibitem{LafargeEstevePRB1993}
P.~Lafarge and D.~Esteve, Phys.~Rev.~B {\bf 48}, 14309
(1993).

\bibitem{Thielmann05}
A.~Thielmann, M.~H.~Hettler, J.~K\"onig, and G.~Sch\"on,
Phys.~Rev.~Lett.~{\bf 95}, 146806 (2005);
A.~Braggio, J.~K\"onig, and R.~Fazio, Phys.~Rev.~Lett.~{\bf 96}, 026805 (2006).

\bibitem{footnote1}
We restrict our formulation here to geometries without direct
junctions between two leads and each lead is connected to one
island only.

\bibitem{vanderWielReview2003}
W.~G.~van der Wiel, S.~De Franceschi, J.~M.~Elzerman,
T.~Fujisawa, S.~Tarucha, and L.~P.~Kouwenhoven,
Rev.~Mod.~Phys.~{\bf 75}, 1 (2003).

\bibitem{footnote2}
The charge
state can be treated as independent from the fermionic degrees of
freedom connected to field operators $a_{l_j\kappa\nu},\;
a^{\dagger}_{l_j\kappa\nu},\; c_{m_j\lambda\nu},\;
c^{\dagger}_{m_j\lambda\nu}$ as the total number of electrons on
any island is large.

\bibitem{SchoellerCuracao1997}
H.~Schoeller, in {\it Mesoscopic Electron Transport}, edited by L.~L.
Sohn, L.~P.~Kouwenhoven, and G.~Sch{\"o}n (Kluwer Academic Press,
The Netherlands, 1997).

\bibitem{footnote4}
The line direction is to be understood as taken with respect to
the convention for the direction of a junction as chosen
above.

\bibitem{footnote5}
Charge states, neighboring the chosen subspace
might appear in these intermediate step, which charging energies
hence are also needed.

\bibitem{footnote6}
According to this systematic expansion the second-order
current contribution includes not only higher-order transition
rates ($P^{(0)}_{\chi'} \Sigma^{(2)}_{\chi',\chi}$) but also
higher-order corrections to the stationary probability
$P^{(1)}_{\chi'} \Sigma^{(1)}_{\chi',\chi}$.

\bibitem{Pekola_thermometry}
M.~Meschke, J.~P.~Pekola, F.~Gay, R.~E.~Rapp, and H.~Godfrin, 
J.~Low Temp.~Phys.~{\bf 134}, 1119 (2004).

\bibitem{SCtwoislands}
D.~C.~Dixon, C.~P.~Heij, P.~Hadley, and J.~E.~Mooij,
J.~Low Temp.~Phys.~{\bf 118}, 325 (2000). 

\bibitem{PofEtheory}
G.-L.~Ingold and Yu.~V.~Nazarov, in {\it Single Charge Tunneling} (Ref.~\onlinecite{GrabertDevoretBook92}), p.~21. 

\bibitem{AguadoQuantumNoisePRL2000}
R.~Aguado and L.~P.~Kouwenhoven, Phys.~Rev.~Lett.~{\bf 84},
1986 (2000).

\bibitem{DelftProceedings2003}
G.~Johansson, P.~Delsing, K.~Bladh, D.~Gunnarsson, T.~Duty, A.
K{\"a}ck, G.~Wendin, and A.~Aassime, in {\em Quantum Noise in
Mesoscopic Physics}, edited by Y.~V.~Nazarov (Kluwer Academic
Publishers, The Netherlands, 2003).

\bibitem{JohanssonSETquantumNoisePRL2002}
G.~Johansson, A.~K\"ack, and G.~Wendin, Phys.~Rev.~Lett.~{\bf
88}, 046802 (2002).

\bibitem{KackSETquantumNoisePRB2003}
A.~K\"ack, G.~Wendin, and G.~Johansson, Phys.~Rev.~B {\bf 67},
035301 (2003).

\bibitem{footnote7}
In principle the relaxation rates $\Gamma_{10}^{D\pm}$ will also be 
modified by the coupling to the
generator. We neglect this modification since it is of order $R_K/R_D$ 
smaller than the first order rates.


\bibitem{NazarovQuantumNoiseDetectionPRL2004}
J.~Tobiska and Yu.~V.~Nazarov, Phys.~Rev.~Lett.~{\bf 93}, 106801
(2004).

\bibitem{SoninNoiseDetectionPRB2004}
E.~B.~Sonin, Phys.~Rev.~B {\bf 70}, 140506(R) (2004).

\bibitem{HakonenNoiseDetectionPRL2004}
R.~K.~Lindell, J.~Delahaye, M.~A.~Sillanp{\"a}{\"a}, T.~T.~Heikkil{\"a}, E.~B.~Sonin, 
and P.~J.~Hakonen, Phys.~Rev.~Lett.~{\bf 93}, 197002 (2004).

\bibitem{PekolaNoiseDetectionPRL2004}
J.~P.~Pekola, Phys.~Rev.~Lett.~{\bf 93}, 206601 (2004).

\bibitem{HeikkilaQuantumNoiseDetectionPRL2004}
T.~T.~Heikkil\"a, P.~Virtanen, G.~Johansson, and F.~K.~Wilhelm,
Phys.~Rev.~Lett.~{\bf 93}, 247005 (2004).

\bibitem{GrabertPRL05}
J.~Ankerhold and H.~ Grabert, Phys.~Rev.~Lett.~{\bf 95}, 186601 (2005).

\bibitem{HartmannNazarov}
E.~Onac, F.~Balestro, L.~H.~W.~van Beveren, U.~Hartmann, Yu.~V.~Nazarov, and L.~P.~Kouwenhoven,
Phys.~Rev.~Lett.~{\bf 96}, 176601 (2006).

\bibitem{ClerkQubitSpectrometer2003}
R.~J.~Schoelkopf, A.~A.~Clerk, S.~M.~Girvin, K.~W.~Lehnert, and M.
H.~Devoret, in {\em Quantum Noise in Mesoscopic Physics}, edited by
Yu.~V.~Nazarov (Kluwer Academic Publishers, The Netherlands, 2003).

\bibitem{DeblockQuantumNoiseDetectionScience2003}
R.~Deblock, E.~Onac, L.~Gurevich, L.~P.~Kouwenhoven, Science {\bf
301}, 203 (2003).

\bibitem{footnote8}
For completeness, we note that the analyzed term, endangering convergence of the series, is not necessarily the largest term, as the principal value parts from the denominators $(\omega -\Delta +i\eta)^{-(n+1)}$ may gain importance. Consider, e.g., standard cotunneling deep in the cotunneling regime, i.e., in the exponentially suppressed region of the rate  $\left| \Delta -\mu_L \right| > \left| \sigma(\Delta)\right|,\,k_BT$. There second-order terms (stemming from those principal value parts) being algebraically suppressed dominate over exponentially suppressed first-order contributions. Nonetheless the convergence argument from above is fulfilled and, indeed, higher than second-order contributions are sufficiently small.

\end{thebibliography}
\end{document}